\documentclass[aps,prl,reprint,superscriptaddress,nofootinbib,longbibliography]{revtex4-2}

\usepackage{amsmath,amssymb,bm,mathtools}
\usepackage{graphicx}
\usepackage{microtype}
\usepackage{xcolor}
\usepackage[colorlinks=true,citecolor=blue,urlcolor=blue,linkcolor=blue]{hyperref}
\usepackage{booktabs}
\usepackage{array}

\newcommand{\Tr}{\operatorname{Tr}}
\newcommand{\arcosh}{\operatorname{arcosh}}
\newcommand{\cA}{\mathcal A}
\newcommand{\cB}{\mathcal B}

\newcommand{\bOne}{\mathbf 1}
\newcommand{\mr}{\mathrm}
\newcommand{\mout}{\mathbf m}
\newcommand{\nout}{\mathbf n}
\newcommand{\ket}[1]{|#1\rangle}
\newcommand{\bra}[1]{\langle #1|}
\newcommand{\hcyl}{\mathrm{hcyl}}
\newcommand{\spec}{\operatorname{spec}}
\newcommand{\Span}{\operatorname{span}}
\newcommand{\diag}{\operatorname{diag}}

\begin{document}

\title{Universal Frame Potential Hierarchy in Critical Projected Ensembles}

\author{Hui-Huang Chen}
\email{chenhh@jxnu.edu.cn}
\affiliation{School of Physics, Jiangxi Normal University, Nanchang 330022, China}

\date{\today}

\begin{abstract}
Local measurements transform a many body wavefunction into a statistical ensemble of conditional quantum states. In chaotic systems, the higher moments of such ensembles diagnose the emergence of quantum state randomness, but their structure at equilibrium quantum criticality is largely unknown. Here we show that the projected ensemble of a Tomonaga-Luttinger liquid exhibits a universal nonlinear hierarchy of state overlap moments. Remarkably, the leading scaling exponents are independent of the continuously varying Luttinger parameter. Replica boundary conformal field theory reveals a geometric origin: outcome locking combines the active replica swaps into a single collective rotated sector, while intersecting compact replica branes eliminate the leading interaction dependent zero mode contribution. Matrix product state calculations across the interacting XXZ critical phase and direct free fermion calculations independently confirm the interaction independence and nonlinear hierarchy.
\end{abstract}

\maketitle

\textit{Introduction.--}
Measurements do more than extract information from a quantum many-body state: they convert a single wavefunction into a statistical ensemble of conditional states. Completely measuring a subsystem $C$ produces conditional pure states on its complement $U$, weighted by their Born probabilities, forming the \emph{projected ensemble} \cite{HoChoi2022,Cotler2023,IppolitiHo2023}. In quantum chaotic dynamics, the higher moments of this ensemble provide a sensitive probe of emergent randomness. Projected ensembles can approach Haar-random state designs in suitable settings, extending thermalization from low order observables to increasingly high moments of the conditional state distribution \cite{IppolitiHo2022,ClaeysLamacraft2022,Choi2023Nature,Lucas2023,Mark2024,ChanDeLuca2024,YuHoKos2025,ChangEtAl2025,Yan2026}. This raises a complementary question for quantum critical states: can the higher order statistics generated by measurement retain universal information about the underlying critical theory?

Critical ground states provide a natural setting in which to address this question. They are scale invariant yet strongly structured, and local measurements are known to preserve universal conformal information in quantities ranging from Shannon type observables to post-measurement entanglement \cite{Stephan2009,Zaletel2011,Stephan2011,AlcarazRajabpour2014,Rajabpour2015,Rajabpour2016,NajafiRajabpour2016}. More recently, measurements of critical states have been studied from the perspectives of measurement-altered criticality, boundary transitions and imperfect teleportation \cite{Garratt2023,Weinstein2023,Yang2023,Murciano2023,LeeJianXu2023,SunYaoJian2023,Cheng2024,KhannaMurcianoVasseur2026,KumarPatilLudwigVasseur2026}, while replica and boundary conformal field theory (BCFT) methods have been developed for measurement-induced entanglement and entanglement swapping \cite{HoshinoOshikawaAshida2025,Khanna:2025mrz,KhannaVasseur2026}. Measurement conditioned ensembles have also begun to be analysed directly in conformal field theory (CFT) \cite{MilekhinMurciano2025}. Most of these works characterize observables within individual conditional states or their measurement averages; frame potentials instead probe the geometry of the projected ensemble itself through overlaps between independently sampled conditional states. Yet these higher overlap moments remain largely unexplored at equilibrium criticality. In particular, it is not known whether they obey universal conformal scaling or whether, in a Tomonaga–Luttinger liquid (TLL), their leading exponents retain the continuously varying interaction parameter.

Here we show that the higher overlap moments of a critical projected ensemble form a universal nonlinear hierarchy. We consider a TLL on a ring and projectively measure the complement $C$ of a single interval $U$ in the local charge basis. For two independent Born distributed measurement outcomes $\mout,\nout$, with corresponding normalized conditional states $|\psi_{\mout}\rangle,|\psi_{\nout}\rangle$, we characterize the ensemble through the frame potentials $F_k=\mathbb E_{\mout,\nout}|\langle\psi_{\mout}|\psi_{\nout}\rangle|^{2k}$. For an interval of length $l$ on a ring of circumference $L$, define the dimensionless chord distance $D_U=(L/\pi\epsilon)\sin(\pi l/L)$, where $\epsilon$ is a short distance cutoff. We find
\[
F_k(U)\sim \cA_k D_U^{-\Gamma_k},\qquad
\Gamma_k=\frac14+\frac{1}{\pi^2}\arcosh^2\!\sqrt{k}.
\]
Here $\cA_k$ is an $l$-independent nonuniversal amplitude. The hierarchy is nonlinear in $k$, yet its leading exponent is independent of the Luttinger parameter. Thus the projected ensemble retains a universal higher order structure that is not reducible to its first moment.

The mechanism is geometric. An auxiliary Born replica construction distinguishes partial permutation sewing on $U$ from \emph{outcome locking} on its measured complement $C$. After folding, the $k$ active swaps collapse into a single collective rotated sector in replica space, whose oscillator twist produces the nonlinear $k$ dependence. In the compact target space the two replica branes intersect, so the neutral compact ground state has zero weight and the Luttinger parameter enters only amplitudes and excited sectors. We test these two consequences separately: interacting XXZ data across the TLL phase probe the predicted Luttinger parameter independence, while a free fermion calculation tests the nonlinear hierarchy together with its conformal distance dependence.

\begin{figure}[t]
    \includegraphics[width=\columnwidth]{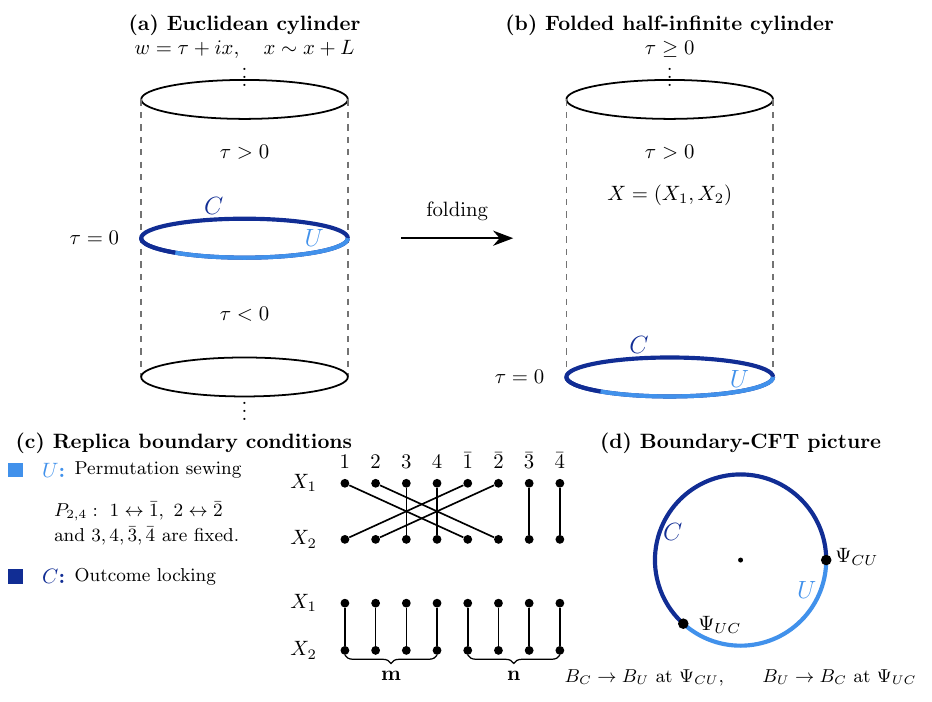}
    \caption{Replica geometry for a single unmeasured interval $U$. (a) The seam at $\tau=0$ carries different gluings on $U$ and its measured complement $C$. (b) Folding turns the seam into a boundary for doubled fields. (c) $U$ carries partial permutation sewing, while outcome locking on $C$ forces replicas in each group to share one measurement record. (d) Mapping to the disk produces boundary condition changing (BCC) insertions $\Psi_{CU}$ and $\Psi_{UC}$ at the endpoints of $U$.}
    \label{fig:folding}
\end{figure}

\textit{Replica trick for the frame potential--}
At low energies the TLL is described by a compact boson CFT
\begin{equation}
S=\frac{1}{8\pi K}\int d^2x\,(\partial_\mu\phi)^2,
\qquad
\phi\sim\phi+2\pi,
\label{eq:action}
\end{equation}
where $K$ is the Luttinger parameter \cite{Haldane1981,Giamarchi2003}. The bosonic field $\phi$ is the coarse grained counting field for the conserved $U(1)$ charge, so that a measurement in the local charge basis corresponds, in the scaling limit, to a measurement of the compact field $\phi$ \cite{NajafiRajabpour2016,Khanna:2025mrz,KhannaVasseur2026}.
We consider the ground state $\ket{0}$ on a ring of circumference $L$, and measure the complement $C$ of an interval $U=[x_1,x_2]$, with length $l=x_2-x_1$. 

Let $\mout$ label a complete measurement outcome on $C$. Its unnormalized conditional state is $\sigma_{\mout}={}_C\!\langle\mout|0\rangle\langle0|\mout\rangle_C$, with Born probability $p_{\mout}=\Tr_U\sigma_{\mout}$ and normalized state $\rho_{\mout}=\sigma_{\mout}/p_{\mout}$. The projected ensemble moment operator is $M_k=\sum_{\mout}p_{\mout}\rho_{\mout}^{\otimes k}$, whose purity defines the frame potential $F_k(U)=\Tr M_k^2$. Introducing an independent outcome $\nout$, this becomes
\begin{equation}
F_k(U)=\sum_{\mout,\nout}p_{\mout}p_{\nout}
\bigl[\Tr_U(\rho_{\mout}\rho_{\nout})\bigr]^k.
\label{eq:Fk}
\end{equation}
Since $\rho_{\mout}=\ket{\psi_{\mout}}\bra{\psi_{\mout}}$ is pure, $\Tr_U(\rho_{\mout}\rho_{\nout})=|\langle\psi_{\mout}|\psi_{\nout}\rangle|^2$, so Eq.~\eqref{eq:Fk} is the $2k$-th overlap moment of the projected ensemble. The normalized conditional states in Eq.~\eqref{eq:Fk} introduce inverse powers of Born probabilities. To overcome this problem, we introduce an auxiliary integer $N\ge k$ and define
\begin{equation}
F_{k,N}(U)=\sum_{\mout,\nout}
\bigl[\Tr_U(\sigma_{\mout}\sigma_{\nout})\bigr]^k
p_{\mout}^{N-k}p_{\nout}^{N-k}.
\label{eq:FkN}
\end{equation}
The physical frame potential is recovered by
\begin{equation}
F_k(U)=\lim_{N\to1}F_{k,N}(U).
\label{eq:FkNlimit}
\end{equation}
The replica construction is especially transparent in operator language. Let $P_{\mout}=\ket{\mout}\!\bra{\mout}_C$ be the projector associated with a complete record $\mout$ on $C$, and define the outcome-locking operator $\Delta_C^{(N)}=\sum_{\mout}P_{\mout}^{\otimes N}$. Labeling the two replica groups as $1,\ldots,N$ and $\bar1,\ldots,\bar N$, let $S_U^{(k)}=\prod_{a=1}^{k}\mathrm{SWAP}_U(a,\bar a)$ swap the first $k$ replicas between the two groups. Then the auxiliary frame potential in Eq.~\eqref{eq:FkN} can be written as
\begin{equation}
F_{k,N}(U)=\Tr\!\left[\rho_0^{\otimes 2N}
\bigl(\Delta_C^{(N)}\otimes\Delta_C^{(N)}\bigr)S_U^{(k)}\right],
\label{eq:operatorrep}
\end{equation}
where $\rho_0=\ket{0}\!\bra{0}$ is the density matrix of the ground state.
In this representation, $S_U^{(k)}$ gives permutation sewing on $U$, whereas $\Delta_C^{(N)}$ locks the replicas within each group to a common measurement outcome on $C$.

\textit{Folded boundary CFT--}
Permutation sewing and outcome locking are distinct boundary operations. On $U$, $S_U^{(k)}$ gives ordinary permutation sewing. Denote $\phi_a^{\pm}$ and $\bar\phi_a^{\pm}$ as the upper and lower bank values at the $\tau=0$ slice in the unbarred and barred replica groups, respectively. For the first $k$ replica pairs,
\begin{equation}
\phi_a^+=\bar\phi_a^-,\qquad \bar\phi_a^+=\phi_a^-,\qquad a\le k,
\label{eq:crosssew}
\end{equation}
while the remaining pairs are sewn within the same replica group. Equivalently, the $2N$ replica labels are acted on by a partial permutation $P_{k,N}$ exchanging $a\leftrightarrow\bar a$ for $a\le k$ and fixing all other labels.

On $C$, by contrast, the outcome locking projector has the continuum kernel
\begin{equation}
\int \mathcal D\mout\,
\prod_{a=1}^{N}\delta[\phi_a^+-\mout]\,\delta[\phi_a^--\mout],
\label{eq:lockingkernel}
\end{equation}
with an independent barred copy involving $\nout$. Integrating over the common records locks all replicas in each group to one common boundary field rather than pairing upper and lower banks. Thus one center of mass direction per group remains Neumann, while all replica relative directions are Dirichlet. Figure~\ref{fig:folding}(c) illustrates the contrast between partial permutation sewing on $U$ and outcome locking on $C$.

Now fold the lower half cylinder onto the upper half cylinder as in Fig.~\ref{fig:folding}. The seam becomes a physical boundary of a theory of $4N$ real bosons. Denote the folded boundary conditions on $U$ and $C$ by $B_U$ and $B_C$, respectively, and their Neumann tangent spaces by $V_U$
and $V_C$. For $B_U$, the Neumann tangent space is the graph of the partial permutation,
\begin{equation}
V_U=\left\{\frac{1}{\sqrt2}(P_{k,N}v,v):v\in\mathbb R^{2N}\right\},
\label{eq:VU}
\end{equation}
For $B_C$, only two uniform directions remain Neumann: $u_1=N^{-1/2}(\bOne_N,\bm{0}_N)$ and $u_2=N^{-1/2}(\bm{0}_N,\bOne_N)$. Here $\bOne_d$ and $\bm{0}_d$ denote the $d$-component all ones and zero vectors. With $c_\alpha=(u_\alpha,u_\alpha)/\sqrt2$ for $\alpha=1,2$, the folded Neumann space is
\begin{equation}
V_C=\operatorname{span}\{c_1,c_2\}.
\label{eq:VC}
\end{equation}
These two linear boundary conditions are conformal. Their junction at each endpoint of $U$ defines boundary condition changing (BCC) operators $\Psi_{CU}$ and $\Psi_{UC}$ \cite{Cardy1984,Cardy1989}. Conformal invariance fixes their two-point function on the folded half cylinder to
\begin{equation}
F_{k,N}(U)\propto
\left\langle\Psi_{CU}(x_1)\Psi_{UC}(x_2)\right\rangle
\sim D_U^{-2h_{k,N}},
\label{eq:BCC2pt}
\end{equation}
where $h_{k,N}$ is the lowest scaling dimension in the mixed $B_C$--$B_U$ open channel, equivalently the BCC scaling dimension, and $D_U$ is the dimensionless chord distance defined previously. The problem therefore reduces to determining $h_{k,N}$.

\textit{BCC spectrum--}
For free bosons, the mixed boundary problem is fixed by the relative geometry of the two Neumann subspaces \cite{Cardy1989,GaberdielRecknagel2001}. Their principal angle decomposition \cite{BjorckGolub1973} yields independent Neumann--Neumann (NN), Dirichlet--Dirichlet (DD), Neumann--Dirichlet (ND), and rotated sectors, the latter being the standard branes at an angle problem \cite{BerkoozDouglasLeigh1996,Pesando2014}.

To determine $h_{k,N}$ we compare the projectors onto $V_C$ and $V_U$. Restricting the projector onto $V_U$ to $V_C$ gives squared principal angle cosines $\lambda_+=1$ and $\lambda_-=1-k/N$. Hence there is one common Neumann direction and a single nontrivial principal angle,
\begin{equation}
\sin^2\theta_{k,N}=\frac{k}{N},
\qquad 0\le \theta_{k,N}\le\frac{\pi}{2}.
\label{eq:theta}
\end{equation}
This angle is collective: the $k$ active swaps do not generate $k$ independent boundary defects, but combine through outcome locking into one rotated two-plane in replica space. The remaining directions consist of $2N-1$ common Dirichlet directions and $2N-2$ real Neumann--Dirichlet mismatch directions. Writing the fractional twist as $\nu_{k,N}\equiv\theta_{k,N}/\pi$, the product of the two boundary reflection matrices contains $2N-2$ eigenvalues $-1$ and one conjugate pair $e^{\pm 2\pi i\nu_{k,N}}$. The fractional moding of this single collective complex boson is the origin of the nonlinear dependence on the moment order $k$.

Each real ND boson contributes $1/16$ to the BCC conformal weight, while the complex boson twisted by $\nu_{k,N}$ contributes $\nu_{k,N}(1-\nu_{k,N})/2$ \cite{Cardy1984,BerkoozDouglasLeigh1996,Pesando2014}. The oscillator contribution is thus
\begin{equation}
h^{\mr{osc}}_{k,N}=\frac{N-1}{8}+\frac12\nu_{k,N}(1-\nu_{k,N}).
\label{eq:hosc}
\end{equation}
The mixed-channel BCC dimension separates as $h_{k,N}=h^{\mr{osc}}_{k,N}+h^{(0)}_{k,N}$, so it remains to determine the compact contribution $h^{(0)}_{k,N}$.

Folding turns the target space into a $4N$-torus. We denote the corresponding compact target space branes by $\cB_U$ and $\cB_C$. With $S^1=\mathbb R/(2\pi\mathbb Z)$,
\begin{align}
\cB_U&=\{(P_{k,N}v,v):v\in (S^1)^{2N}\},\nonumber\\
\cB_C&=\{(\alpha\bOne_N,\beta\bOne_N;\alpha\bOne_N,\beta\bOne_N):\alpha,\beta\in S^1\}.
\label{eq:branes}
\end{align}
For any $k>0$ they intersect along the diagonal circle
\begin{equation}
\cB_C\cap\cB_U=\{\phi_0\,\bOne_{4N}:\phi_0\in S^1\}.
\label{eq:intersection}
\end{equation}
Thus the mixed $B_C$--$B_U$ channel always admits a constant classical configuration with zero action, implying
\begin{equation}
h^{(0)}_{k,N}=0.
\label{eq:hzero}
\end{equation}
Geometrically, the two replica branes intersect, so the mixed channel pays no classical stretching or winding cost. This is the reason the Luttinger parameter $K$ cannot enter the leading single-interval exponent: compact momentum and winding excitations depend on $K$, but the neutral compact ground state remains at zero weight. For the leading single interval scaling considered here, only the neutral ground sector in Eq.~\eqref{eq:hzero} is required.

\textit{Replica continuation and universal hierarchy--}
The BCFT construction above is controlled at positive integer $N\ge k$. In that domain, the auxiliary frame potential has the scaling form
\begin{equation}
F_{k,N}(U)=\cA_{k,N}D_U^{-\Gamma_{k,N}},
\label{eq:integer-scaling}
\end{equation}
where $\cA_{k,N}$ is an $l$-independent nonuniversal amplitude. Combining the oscillator ground state energy with the compact result $h^{(0)}_{k,N}=0$ gives the exponent
\begin{equation}
\Gamma_{k,N}\equiv2h_{k,N}=\frac{N-1}{4}+\nu_{k,N}(1-\nu_{k,N}).
\label{eq:GammaNk}
\end{equation}
Compactification modifies amplitudes and excited sectors but not the leading single interval exponent.

The physical frame potential requires $N\to1$. For $k>1$, this lies outside the integer domain $N\ge k$, so we analytically continue the mixed-channel ground-state branch selected by the BCFT. Its relative monodromy obeys the invariant relation
\begin{equation}
\cos(2\pi\nu_{k,N})=1-\frac{2k}{N}.
\label{eq:monodromy-invariant}
\end{equation}
Continuity of this branch through $N=k$ gives, for $N<k$,
\begin{equation}
\nu_{k,N}=\frac12\pm\frac{i}{\pi}\arcosh\sqrt{\frac{k}{N}},
\label{eq:continuation}
\end{equation}
The two signs give complex conjugate values of $\nu_{k,N}$ and the same real combination $\nu_{k,N}(1-\nu_{k,N})$. Continuing the same branch to $N=1$ yields
\begin{equation}
\Gamma_k=\lim_{N\to1}\Gamma_{k,N}=\frac14+\frac{1}{\pi^2}\arcosh^2\!\sqrt{k}.
\label{eq:Gamma}
\end{equation}
Thus,
\begin{equation}
F_k(U)= \cA_k
\left[\frac{L}{\pi\epsilon}\sin\!\left(\frac{\pi l}{L}\right)\right]^{-\frac14-\pi^{-2}\arcosh^2\sqrt{k}}.
\label{eq:main-result}
\end{equation}
The case $k=1$ requires no nontrivial continuation and provides an exact check:
\begin{equation}
F_1(U)=\Tr\!\left(\sum_{\mout}p_{\mout}\rho_{\mout}\right)^2
=\Tr\rho_U^2\sim D_U^{-1/4},
\label{eq:k1}
\end{equation}
where $\rho_U=\Tr_C\rho_0$ is the ground-state reduced density matrix on $U$, consistent with the standard second R\'enyi purity of a $c=1$ CFT \cite{CalabreseCardy2004}. The resulting hierarchy also satisfies the general moment constraints of pure-state frame potentials, including monotonicity and log-convexity; see Supplemental Material \cite{SupplementalMaterial}.

\begin{figure*}[t]
    \centering
    \begin{minipage}[t]{0.55\textwidth}
        \centering
        \textbf{(a)}\\[-0.5ex]
        \includegraphics[width=\linewidth]{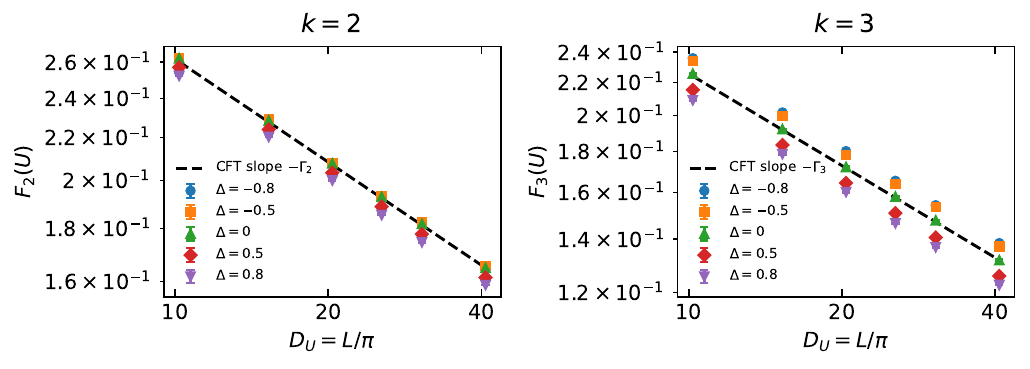}
    \end{minipage}\hfill
    \begin{minipage}[t]{0.41\textwidth}
        \centering
        \textbf{(b)}\\[-0.5ex]
        \includegraphics[width=\linewidth]{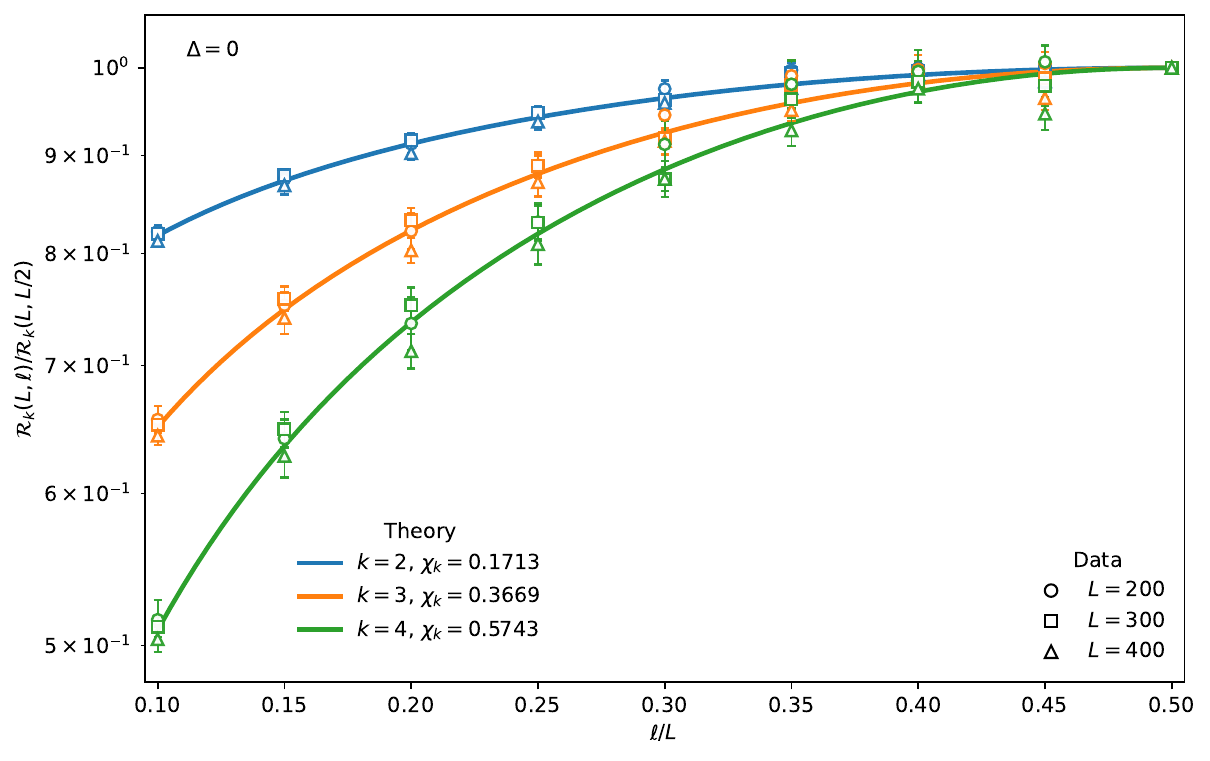}
    \end{minipage}
    \caption{Two complementary tests of distinct predictions. (a) Interacting periodic XXZ data for $k=2,3$ and five anisotropies test the predicted independence of $\Gamma_k$ from the Luttinger parameter $K$ across the TLL phase. Dashed lines have the parameter-free CFT slopes $-\Gamma_k$ from Eq.~\eqref{eq:Gamma} and are anchored only in vertical position to the largest size $\Delta=0$ datum. (b) The free fermion calculation tests the nonlinear hierarchy and conformal distance dependence for $k=2,3,4$ and $L=200,300,400$. Solid curves are the parameter-free prediction in Eq.~\eqref{eq:shape}. Statistical error bars are shown in both panels.}
    \label{fig:tests}
\end{figure*}

\textit{Numerical test--}
We perform two independent calculations directly in the physical projected ensemble, neither of which relies on the replica continuation. The first uses MPS representations of interacting XXZ ground states and varies the anisotropy to test both the predicted exponents and their independence of the Luttinger parameter $K$. The second exploits the free fermion structure at $\Delta=0$ to reach larger sizes, higher moments, and arbitrary interval fractions, providing an independent test of the nonlinear hierarchy and its conformal distance dependence.

For the first test, consider the periodic spin-$1/2$ XXZ chain (with the transverse exchange set to unity)
\begin{equation}
H=\sum_{j=1}^{L}\left(S_j^xS_{j+1}^x+S_j^yS_{j+1}^y+\Delta S_j^zS_{j+1}^z\right),
\label{eq:xxz}
\end{equation}
where $S_j^{\alpha}$ are spin-$1/2$ operators, $\Delta$ is the exchange anisotropy, and $S_{L+1}\equiv S_1$. At zero magnetization, the Bethe ansatz relation gives $K(\Delta)=\pi/[2(\pi-\arccos\Delta)]$ for $-1<\Delta\le1$, with $K=1$ at the free fermion point $\Delta=0$ \cite{Haldane1981,Giamarchi2003}. We obtain matrix product state approximations to the ground state using density matrix renormalization group (DMRG) \cite{White1992,White1993} implemented with the ITensor library \cite{ITensor2022}. The complement of the half chain is projectively measured in the local $S^z$ basis, and measurement outcomes are sampled according to the Born rule. For two independent outcomes $\mout,\nout$, the conditional state overlap $X_{\mout\nout}=|\langle\psi_{\mout}|\psi_{\nout}\rangle|^2$ gives $F_k=\mathbb E[X_{\mout\nout}^k]$; the exact $F_1=\Tr\rho_U^2$ from the Schmidt spectrum is used as a control variate. Figure~\ref{fig:tests}(a) shows $k=2,3$ data for $\Delta=-0.8,-0.5,0,0.5,0.8$ and $L\le128$, all within the critical TLL phase $-1<\Delta<1$. Across these representative interacting TLL points, the data follow the same parameter free CFT slopes $-\Gamma_k$; changing $\Delta$ primarily shifts the nonuniversal amplitude.

For the second test, at $\Delta=0$, the XXZ chain maps to free fermions, for which Gaussianity allows an independent Slater determinant evaluation at substantially larger system sizes; see Supplemental Material \cite{SupplementalMaterial}. Normalizing $\mathcal R_k\equiv F_k/F_1^k$ by its half chain value removes the nonuniversal amplitude and gives the parameter-free shape prediction
\begin{equation}
\frac{\mathcal R_k(L,l)}{\mathcal R_k(L,L/2)}
\sim \left[\sin\!\left(\frac{\pi l}{L}\right)\right]^{\chi_k},
\label{eq:shape}
\end{equation}
where $\chi_k=k\Gamma_1-\Gamma_k$. Figure~\ref{fig:tests}(b) shows $k=2,3,4$ data for $L=200,300,400$. The three sizes collapse onto the predicted curves over the accessible interval range, testing the nonlinear hierarchy and its conformal distance dependence.

\textit{Discussion and outlook--}
The main implication is that projected ensembles at equilibrium criticality possess universal structure well beyond their first moment. In a TLL, higher overlap moments form a nonlinear conformal hierarchy rather than simple powers of the purity, providing a critical counterpart to the moment diagnostics used in deep thermalization. The persistence of the same leading exponents across continuously varying $K$ along the TLL fixed line makes this hierarchy a particularly sharp observable of the measurement geometry.

A natural next step is to move beyond a single interval. With multiple unmeasured intervals, BCC correlators acquire conformal cross ratios and the compact sector admits nontrivial winding configurations, so the answer is no longer fixed by one local BCC dimension. A second direction is to resolve the projected ensemble by the global $U(1)$ symmetry. Charge-resolved projected ensembles should connect the present boundary construction with charged moments and symmetry-resolved entanglement \cite{ChangEtAl2025,GoldsteinSela2018,BonsignoriRuggieroCalabrese2019,KusukiMurcianoOoguriPal2023}, and provide a controlled setting in which compact zero modes become directly observable. More broadly, holographic measurement setups raise the question of whether these higher overlap moments and the mixed-channel BCC spectrum admit a semiclassical bulk interpretation \cite{AntoniniEtAl2022,AntoniniEtAl2023}.

\textit{Acknowledgments--} This work was supported by the National Natural Science Foundation of China, Grant No.
12005081 and No. 12465014.

\clearpage
\onecolumngrid

\begin{center}
{\large \textbf{Supplemental Material for\\
``Universal Frame Potential Hierarchy in Critical Projected Ensembles''}}

\vspace{0.5cm}

Hui-Huang Chen
\end{center}

\setcounter{equation}{0}
\setcounter{figure}{0}
\setcounter{table}{0}
\setcounter{section}{0}

\renewcommand{\theequation}{S\arabic{equation}}
\renewcommand{\thetable}{S\Roman{table}}
\renewcommand{\thesection}{S\arabic{section}}

This Supplemental Material provides the technical derivations and numerical details underlying the main text.

\begin{center}
\textbf{CONTENTS}
\end{center}
\noindent  Replica formulation and folded boundary CFT \dotfill \pageref{sec:S1}\\
 Collective boundary geometry and oscillator spectrum \dotfill \pageref{sec:S2}\\
 Compact ground sector and $K$ dependence\dotfill \pageref{sec:S3}\\
 Frame potential inequalities \dotfill \pageref{sec:S4}\\
 Numerical methods \dotfill \pageref{sec:S5}
\vspace{0.8em}

\section{Replica formulation and folded boundary CFT}
\label{sec:S1}

\subsection{Auxiliary replica identity}

We use the compact-boson normalization,
\begin{equation}
S=\frac{1}{8\pi K}\int d^2x\,(\partial_\mu\phi)^2,
\qquad \phi\sim\phi+2\pi,
\label{eq:S-action}
\end{equation}
where $K$ is the Luttinger parameter \cite{Haldane1981,Giamarchi2003}. Consider the ground state $\rho_0=\ket{0}\!\bra{0}$ on a circle of circumference $L$. The unmeasured interval is $U=[x_1,x_2]$, with length $l=x_2-x_1$, and the measured complement is $C=U^c$. 

Let $\mout$ denote a complete rank-one projective-measurement outcome on $C$ and $P_{\mout}=\ket{\mout}\!\bra{\mout}_C$ its projector. In the notation of the main text, the unnormalized conditional state on $U$ is
$\sigma_{\mout}={}_C\!\langle\mout|\rho_0|\mout\rangle_C$, its Born probability is $p_{\mout}=\Tr_U\sigma_{\mout}$, and $\rho_{\mout}=\sigma_{\mout}/p_{\mout}$ is the normalized conditional state. Introducing an independent outcome $\nout$, the $k$th projected-ensemble frame potential is
\begin{equation}
F_k(U)=\sum_{\mout,\nout}p_{\mout}p_{\nout}
\bigl[\Tr_U(\rho_{\mout}\rho_{\nout})\bigr]^k.
\label{eq:S-Fk}
\end{equation}
For the present rank-one protocol $\rho_{\mout}$ and $\rho_{\nout}$ are pure, so the bracket is $|\langle\psi_{\mout}|\psi_{\nout}\rangle|^2$.

The normalization of the conditional states is the only obstruction to a direct integer-replica path integral. Using
$\Tr_U(\rho_{\mout}\rho_{\nout})=\Tr_U(\sigma_{\mout}\sigma_{\nout})/(p_{\mout}p_{\nout})$, introduce an auxiliary integer $N\ge k$ and define
\begin{equation}
F_{k,N}(U)=\sum_{\mout,\nout}
\bigl[\Tr_U(\sigma_{\mout}\sigma_{\nout})\bigr]^k
p_{\mout}^{N-k}p_{\nout}^{N-k}.
\label{eq:S-FkN}
\end{equation}
The physical frame potential is recovered by the replica continuation
\begin{equation}
F_k(U)=\lim_{N\to1}F_{k,N}(U).
\label{eq:S-limit}
\end{equation}
The construction up to this point is algebraic. The BCFT analysis below is first performed at positive integers $N\ge k$, where Eq.~\eqref{eq:S-FkN} is represented by an ordinary replicated Euclidean path integral.

To expose its connectivity, define the outcome-locking operator
$\Delta_C^{(N)}=\sum_{\mout}P_{\mout}^{\otimes N}$. Label the two groups of $N$ replicas by $1,\ldots,N$ and $\bar1,\ldots,\bar N$, and let
$S_U^{(k)}=\prod_{a=1}^{k}\mathrm{SWAP}_U(a,\bar a)$ exchange the first $k$ replicas between the groups on $U$. Then
\begin{equation}
F_{k,N}(U)=\Tr\!\left[
\rho_0^{\otimes 2N}
\bigl(\Delta_C^{(N)}\otimes\Delta_C^{(N)}\bigr)
S_U^{(k)}\right].
\label{eq:S-operator}
\end{equation}
A direct contraction verifies this identity. Expanding the two locking operators and fixing $\mout,\nout$ gives
\begin{equation}
\Tr\!\left[
\rho_0^{\otimes 2N}
\bigl(P_{\mout}^{\otimes N}\otimes P_{\nout}^{\otimes N}\bigr)
S_U^{(k)}\right]
=
\bigl[\Tr_U(\sigma_{\mout}\sigma_{\nout})\bigr]^k
p_{\mout}^{N-k}p_{\nout}^{N-k}.
\label{eq:S-check}
\end{equation}
Each active SWAP closes one unbarred--barred pair into an overlap trace, whereas every spectator replica closes within its own outcome group and produces one Born probability factor. Summing over $\mout,\nout$ recovers Eq.~\eqref{eq:S-FkN}. This form also isolates the two logically different operations that become boundary conditions in the continuum: $S_U^{(k)}$ is an invertible permutation sewing on $U$, while $\Delta_C^{(N)}$ is a projector that locks all replicas in one group to a single measurement record on $C$.

\subsection{Partial permutation on the unmeasured interval}

Collect the $2N$ replica fields before folding into
$\bm\phi=(\phi_1,\ldots,\phi_N;\bar\phi_1,\ldots,\bar\phi_N)^T$. Introduce
$\Pi_k=\diag(\underbrace{1,\ldots,1}_{k},\underbrace{0,\ldots,0}_{N-k})$ and the $2N\times2N$ partial permutation
\begin{equation}
P_{k,N}=
\begin{pmatrix}
I_N-\Pi_k & \Pi_k\\
\Pi_k & I_N-\Pi_k
\end{pmatrix},
\qquad
P_{k,N}^T=P_{k,N},\qquad P_{k,N}^2=I_{2N}.
\label{eq:S-P}
\end{equation}
Thus $P_{k,N}$ exchanges $a\leftrightarrow\bar a$ for $a\le k$ and fixes all labels with $a>k$.
Writing $\bm\phi^\pm(x)=\bm\phi(0^\pm,x)$, the SWAP operator imposes the usual replica sewing on $U$,
\begin{equation}
\bm\phi^+=P_{k,N}\bm\phi^-,
\qquad
\partial_\tau\bm\phi^+=P_{k,N}\partial_\tau\bm\phi^-,
\qquad x\in U.
\label{eq:S-sewing}
\end{equation}
Thus the active replicas are exchanged across the seam while the spectators are sewn to themselves. The same $P_{k,N}$ acts on field values and normal derivatives because the replica theory is a direct product of identical free bosons.

\subsection{Outcome locking on the measured complement}

The measured region has a qualitatively different constraint. Suppressing the barred group for a moment, the continuum kernel of $\Delta_C^{(N)}$ can be written as
\begin{equation}
\int\mathcal D\mout\,
\prod_{a=1}^{N}
\delta[\phi_a^+-\mout]\,\delta[\phi_a^--\mout],
\label{eq:S-lock}
\end{equation}
with an independent barred copy involving $\nout$. For fixed $\mout$, all upper and lower boundary legs in the unbarred group take the same boundary value; the barred group is independently fixed to $\nout$. Compact equalities are understood modulo $2\pi$.

It is important that $\mout(x)$ and $\nout(x)$ are \emph{integrated over}, rather than fixed external Dirichlet profiles. Equality of the replica fields fixes all replica relative combinations, but the common value itself is free to vary. The variational principle makes the resulting Neumann condition explicit. For the unbarred group, an allowed common variation has $\delta\phi_a^+=\delta\phi_a^-=\delta\mout(x)$ for every $a$. The boundary variation of Eq.~\eqref{eq:S-action} is therefore proportional to
\begin{equation}
\delta\mout(x)\sum_{a=1}^N\bigl(\partial_n\phi_a^++\partial_n\phi_a^-\bigr).
\end{equation}
Since $\delta\mout(x)$ is arbitrary after the outcome integral, stationarity requires the normal derivative of the normalized common mode to vanish. Hence one center of mass direction in each outcome group is Neumann, whereas the $N-1$ relative combinations in that group are Dirichlet. The same statement holds for the barred group. Notice that if the outcome profiles were held fixed instead of integrated, the common directions would also be Dirichlet.

\subsection{Folding and the two Neumann tangent spaces}

Fold the lower half-cylinder onto the upper one by defining, for $\tau\ge0$,
$X_1(\tau,x)=\bm\phi(\tau,x)$ and $X_2(\tau,x)=\bm\phi(-\tau,x)$. The folded field
$X=(X_1,X_2)^T$ has $4N$ real components and, globally, takes values in $(S^1)^{4N}$ with $S^1=\mathbb R/(2\pi\mathbb Z)$. Locally in oscillator space we work in the tangent space $\mathbb R^{4N}$.

On $U$, Eq.~\eqref{eq:S-sewing} becomes
\begin{equation}
X_1-P_{k,N}X_2=0,
\qquad
\partial_\tau X_1+P_{k,N}\partial_\tau X_2=0.
\label{eq:S-foldU}
\end{equation}
The first relation fixes the Dirichlet normal combinations, while the second is the corresponding Neumann condition after reversal of the lower half cylinder normal. An allowed boundary displacement satisfies $\delta X_1=P_{k,N}\delta X_2$; hence the Neumann tangent space of the unmeasured boundary condition $B_U$ is the graph
\begin{equation}
V_U=\left\{\frac{1}{\sqrt2}(P_{k,N}v,v):v\in\mathbb R^{2N}\right\},
\qquad \dim V_U=2N.
\label{eq:S-VU}
\end{equation}

For $B_C$, introduce the two normalized uniform replica-label vectors
\begin{equation}
u_1=N^{-1/2}(\bOne_N,\bm0_N),\qquad
u_2=N^{-1/2}(\bm0_N,\bOne_N),
\end{equation}
where $\bOne_d$ and $\bm0_d$ denote the $d$-component all ones and zero vectors. Their normalized folded representatives are $c_\alpha=(u_\alpha,u_\alpha)/\sqrt2$, $\alpha=1,2$. The outcome sums make precisely these two common directions Neumann, so
\begin{equation}
V_C=\Span\{c_1,c_2\},
\qquad \dim V_C=2.
\label{eq:S-VC}
\end{equation}
All $4N-2$ directions orthogonal to $V_C$ are Dirichlet at $B_C$. The strong dimensional asymmetry $\dim V_C=2$ versus $\dim V_U=2N$ is a direct consequence of outcome locking: only two common profiles fluctuate on $C$, whereas all $2N$ replica-label directions fluctuate on $U$ subject to the graph relation.

For a multi-component free boson, a linear conformal boundary condition with Neumann tangent space $V$ can be encoded by the orthogonal reflection
$R_V=2\Pi_V-I$, where $\Pi_V$ is the orthogonal projector onto $V$ \cite{Cardy1989,GaberdielRecknagel2001}. $R_V$ acts as $+1$ on Neumann directions and $-1$ on Dirichlet directions. The relative matrix $R_U R_C$ will therefore contain the complete local oscillator information of the $B_C$--$B_U$ junction.

The boundary condition changes from $B_C$ to $B_U$ at $x_1$ and back from $B_U$ to $B_C$ at $x_2$. The folded partition function therefore has the BCFT representation
\begin{equation}
F_{k,N}(U)\propto
\left\langle\Psi_{CU}(x_1)\Psi_{UC}(x_2)\right\rangle_{\hcyl}\propto D_U^{-2h_{k,N}},
\qquad \Gamma_{k,N}=2h_{k,N}.
\label{eq:S-BCCcorr}
\end{equation}
where  $h_{k,N}$ is the lowest scaling dimension in the mixed $B_C$--$B_U$ open channel.
The remaining BCFT problem is therefore to determine the lowest $B_C$--$B_U$ open channel weight.

\section{Collective boundary geometry and oscillator spectrum}
\label{sec:S2}

\subsection{Principal angle theorem for two Neumann subspaces}

The replica boundary conditions define two Neumann subspaces of different dimensions. Their relative geometry, and hence the mixed oscillator spectrum, is most conveniently characterized by principal angles. Since this construction is less standard in the BCFT literature than ordinary one-dimensional branes at an angle problems \cite{BerkoozDouglasLeigh1996,Pesando2014}, we summarize the required linear algebra results \cite{BjorckGolub1973}. Let $V,W\subset\mathbb R^D$ be Euclidean subspaces with dimensions $p$ and $q$, and let $m=\min(p,q)$. Their principal angles are an ordered set
\begin{equation}
0\le\theta_1\le\cdots\le\theta_m\le\frac\pi2,
\end{equation}
with orthonormal principal vectors $v_i\in V$, $w_i\in W$ that may be chosen so that
$v_i^T w_j=\delta_{ij}\cos\theta_i$. The variational definition successively maximizes the overlap $v^T w$ subject to orthogonality to the previous principal pairs.

For computation, the key statement is the \emph{restricted projector theorem}. If $\Pi_V$ and $\Pi_W$ are the orthogonal projectors, then the self-adjoint positive operator
\begin{equation}
\mathcal B_V\equiv\left.\Pi_V\Pi_W\Pi_V\right|_V:V\to V
\end{equation}
has eigenvalues $\cos^2\theta_i$ (plus additional zeros when $p>q$). If $Q_V$ and $Q_W$ are matrices with orthonormal columns spanning $V$ and $W$, this is equivalently
\begin{equation}
Q_V^T\Pi_WQ_V=(Q_V^TQ_W)(Q_V^TQ_W)^T,
\end{equation}
so the singular values of the overlap matrix $Q_V^TQ_W$ are $\cos\theta_i$. This distinction is useful in practice: \emph{eigenvalues of the restricted projector are $\cos^2\theta_i$, whereas singular values of the basis overlap matrix are $\cos\theta_i$}.

If $0<\theta_i<\pi/2$ and $v_i$ is a normalized eigenvector of $\mathcal B_V$, its partner principal vector is obtained by projection,
\begin{equation}
w_i=\frac{\Pi_Wv_i}{\cos\theta_i},\qquad
\Pi_Wv_i=\cos\theta_i\,w_i,\qquad
\Pi_Vw_i=\cos\theta_i\,v_i.
\end{equation}
The pair spans a two-dimensional principal plane $\mathcal P_i=\Span\{v_i,w_i\}$. With
$n_i=(w_i-\cos\theta_i v_i)/\sin\theta_i$, the basis $(v_i,n_i)$ is orthonormal and
$w_i=\cos\theta_i v_i+\sin\theta_i n_i$. Distinct nontrivial principal planes are mutually orthogonal. A zero principal angle is a common direction in $V\cap W$; a right angle is an orthogonal mismatch. When $p\ne q$, however, there are only $m=\min(p,q)$ principal angles: additional unpaired directions in the larger subspace must be counted separately. This last point will be essential because $\dim V_C=2$ while $\dim V_U=2N$.

Finally, the principal-angle normal form is exactly the one needed for free-boson boundary conditions. On a nontrivial principal plane, the reflections $R_V=2\Pi_V-I$ and $R_W=2\Pi_W-I$ reduce in the basis $(v_i,n_i)$ to reflections about two lines separated by $\theta_i$; their product is a rotation by twice that angle,
\begin{equation}
R_WR_V\big|_{\mathcal P_i}=
\begin{pmatrix}
\cos2\theta_i&-\sin2\theta_i\\
\sin2\theta_i&\cos2\theta_i
\end{pmatrix},
\qquad
\spec(R_WR_V)|_{\mathcal P_i}=\{e^{+2i\theta_i},e^{-2i\theta_i}\}.
\end{equation}
This doubled-angle theorem is the bridge between the Euclidean geometry of Neumann subspaces and the fractional monodromy phases of BCFT currents.

\subsection{Restricted projector for $V_C$ and $V_U$}

We now apply the theorem to the boundary spaces in Eqs.~\eqref{eq:S-VU} and \eqref{eq:S-VC}. An orthonormal embedding of $V_U$ is
\begin{equation}
E_U=\frac{1}{\sqrt2}
\begin{pmatrix}P_{k,N}\\ I_{2N}\end{pmatrix},
\qquad
\Pi_U=E_UE_U^T
=\frac12
\begin{pmatrix}
I_{2N}&P_{k,N}\\
P_{k,N}&I_{2N}
\end{pmatrix}.
\label{eq:S-PiU}
\end{equation}
For $V_C$, set $E_C=(c_1,c_2)$, so $E_C^TE_C=I_2$ and $\Pi_C=E_CE_C^T$. Because $V_C$ is only two-dimensional, the full $4N$-dimensional principal-angle problem reduces exactly to the $2\times2$ matrix
$A=E_C^T\Pi_UE_C$, which represents $\Pi_C\Pi_U\Pi_C|_{V_C}$ in the orthonormal basis $(c_1,c_2)$.

The required replica-label overlaps follow directly from the action of the partial swap:
\begin{equation}
u_1^TP_{k,N}u_1=u_2^TP_{k,N}u_2=\frac{N-k}{N},
\qquad
u_1^TP_{k,N}u_2=u_2^TP_{k,N}u_1=\frac{k}{N}.
\end{equation}
Hence
\begin{equation}
A=
\begin{pmatrix}
1-\dfrac{k}{2N}&\dfrac{k}{2N}\\[4pt]
\dfrac{k}{2N}&1-\dfrac{k}{2N}
\end{pmatrix}.
\label{eq:S-A}
\end{equation}
Its normalized eigenvectors are the symmetric and antisymmetric combinations
$c_+=(c_1+c_2)/\sqrt2$ and $c_-=(c_1-c_2)/\sqrt2$, with eigenvalues
$\lambda_+=1$ and $\lambda_-=1-k/N$. By the restricted-projector theorem these are squared principal-angle cosines. Therefore one principal angle is exactly zero and the other obeys
\begin{equation}
\sin^2\theta_{k,N}=\frac{k}{N},
\qquad
0\le\theta_{k,N}\le\frac\pi2.
\label{eq:S-theta}
\end{equation}
The unit eigenvalue has a simple physical meaning: $\Pi_Uc_+=c_+$, so $c_+$ is a common Neumann direction. It is the simultaneous uniform shift of the two outcome groups. For $0<k<N$, define the partner principal vector
\begin{equation}
w_-\equiv\frac{\Pi_Uc_-}{\cos\theta_{k,N}}\in V_U,
\qquad
c_-^Tw_-=\cos\theta_{k,N},
\end{equation}
and the orthonormal transverse vector
\begin{equation}
n_-\equiv\frac{w_--\cos\theta_{k,N}c_-}{\sin\theta_{k,N}}.
\end{equation}
Then $\mathcal P_{k,N}=\Span\{c_-,w_-\}=\Span\{c_-,n_-\}$ is the unique nontrivial principal plane, with $V_C$ contributing the line $\Span\{c_-\}$ and $V_U$ the line $\Span\{w_-\}$. Thus the $k$ active swaps do \emph{not} create $k$ independent twist planes: outcome locking collapses their effect into a single collective two-plane whose angle depends on the ratio $k/N$.
\begin{figure}[t]
\centering
\includegraphics[width=9cm]{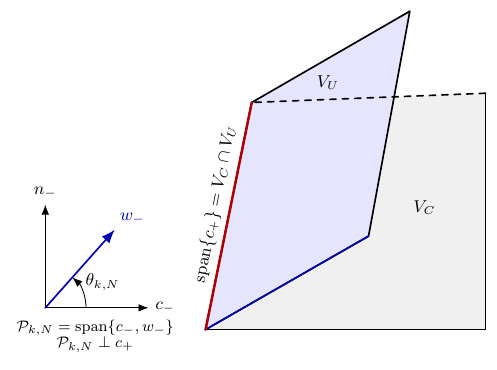}
\caption{Principal-angle geometry of the folded Neumann subspaces for $0<k<N$.
The measured-boundary Neumann space $V_C$ and the unmeasured-boundary Neumann space $V_U$ share the common NN direction $c_+$.
Restricting to the three-dimensional subspace $\operatorname{span}\{c_+,c_-,w_-\}$ makes them appear as two planes meeting along $V_C\cap V_U=\operatorname{span}\{c_+\}$.
The section perpendicular to $c_+$ is the nontrivial principal plane $\mathcal P_{k,N}=\operatorname{span}\{c_-,w_-\}$, where $c_-\in V_C$ and $w_-\in V_U$ meet at the principal angle $\theta_{k,N}$; equivalently, $w_-=\cos\theta_{k,N}\,c_-+\sin\theta_{k,N}\,n_-$.
The additional $2N-2$ directions in $V_U\cap V_C^{\perp}$ are genuine ND/DN mismatch sectors and are not visible in this three-dimensional schematic.}
\label{fig:principal-angle-geometry}
\end{figure}
\subsection{Canonical sector counting}

The principal-angle theorem also clarifies the remaining $4N-2$ directions. For $0<k<N$, there is exactly one common Neumann direction,
$\dim(V_C\cap V_U)=1$. The common Dirichlet sector is
$V_C^\perp\cap V_U^\perp=(V_C+V_U)^\perp$, so
\begin{equation}
\dim(V_C^\perp\cap V_U^\perp)
=4N-\bigl(\dim V_C+\dim V_U-\dim(V_C\cap V_U)\bigr)
=2N-1.
\end{equation}
The two-dimensional principal plane consumes one direction from $V_C$ and one from $V_U$. Since $V_U$ has dimension $2N$, its remaining $2N-2$ unpaired directions lie in $V_U\cap V_C^\perp$ and are Neumann for $B_U$ but Dirichlet for $B_C$. They are genuine ND/DN mismatch sectors; they are \emph{not} extra principal angles, because the standard principal-angle list has only $\min(2,2N)=2$ entries. There are no additional directions in $V_C\cap V_U^\perp$ when $0<k<N$.

Thus the folded tangent space decomposes orthogonally as
\begin{equation}
\mathbb R^{4N}=\mathcal H_{\mr{NN}}\oplus\mathcal H_{\mr{DD}}
\oplus\mathcal H_{\mr{ND}}\oplus\mathcal P_{k,N},
\end{equation}
with dimensions $1$, $2N-1$, $2N-2$, and $2$, respectively. The count
$1+(2N-1)+(2N-2)+2=4N$ provides a useful check. At the endpoint $k=N$, $\theta_{k,N}=\pi/2$ and the principal plane itself becomes two orthogonal mismatch directions, as expected by continuity.

\subsection{Relative monodromy and fractional modes}

Let $R_C=2\Pi_C-I_{4N}$ and $R_U=2\Pi_U-I_{4N}$. The relative monodromy for the BCC operator $\Psi_{CU}$ is $M_{CU}=R_UR_C$, while $M_{UC}=R_CR_U=M_{CU}^{-1}$. Common NN and DD directions contribute eigenvalue $+1$, and each ND mismatch contributes $-1$.

In the principal plane use the orthonormal basis $(c_-,n_-)$, in which the $B_C$ Neumann line is the first axis and the $B_U$ Neumann line is rotated by $\theta_{k,N}$. Then
\begin{equation}
R_C=
\begin{pmatrix}1&0\\0&-1\end{pmatrix},
\qquad
R_U=
\begin{pmatrix}
\cos2\theta_{k,N}&\sin2\theta_{k,N}\\
\sin2\theta_{k,N}&-\cos2\theta_{k,N}
\end{pmatrix},
\label{eq:S-Rs}
\end{equation}
and therefore
\begin{equation}
R_UR_C=
\begin{pmatrix}
\cos2\theta_{k,N}&-\sin2\theta_{k,N}\\
\sin2\theta_{k,N}&\cos2\theta_{k,N}
\end{pmatrix}.
\label{eq:S-Rprod}
\end{equation}
Writing the fractional twist as $\nu_{k,N}=\theta_{k,N}/\pi$, the full relative-monodromy spectrum is
\begin{equation}
\spec M_{CU}=
\left\{+1^{(2N)},\,-1^{(2N-2)},\,
 e^{+2\pi i\nu_{k,N}},\,e^{-2\pi i\nu_{k,N}}\right\}.
\label{eq:S-spec}
\end{equation}
The nonlinear dependence on $k$ is therefore carried by a \emph{single} collective complex boson; the other nontrivial sectors are the $2N-2$ universal half-twists caused by the dimension mismatch of $V_C$ and $V_U$.

The $-1$ sectors therefore have half-integer oscillator moding, while the rotated complex boson has fractional mode shifts $\nu_{k,N}$ and $1-\nu_{k,N}$, as in the standard branes-at-angles construction \cite{BerkoozDouglasLeigh1996,Pesando2014}. Reversing the BCC orientation complex-conjugates the monodromy eigenvalues but leaves the vacuum energy unchanged because it is symmetric under $\nu\leftrightarrow1-\nu$.

\subsection{Oscillator BCC weight}

The BCC vacuum weight is the shift of the oscillator normal-ordering constant relative to an untwisted boundary vacuum. For one real ND boson,
\begin{equation}
h_{\mr{ND}}
=\frac12\left[\zeta\!\left(-1,\frac12\right)-\zeta(-1,1)\right]
=\frac1{16},
\label{eq:S-hND}
\end{equation}
where $\zeta(s,a)$ is the Hurwitz zeta function. Therefore the $2N-2$ mismatch directions contribute $(N-1)/8$. For the complex rotated boson,
\begin{equation}
h_{\mr{rot}}(\nu)
=\frac12\bigl[\zeta(-1,\nu)+\zeta(-1,1-\nu)-2\zeta(-1,1)\bigr]
=\frac12\nu(1-\nu),
\label{eq:S-hrot}
\end{equation}
using $\zeta(-1,a)=-\tfrac12(a^2-a+1/6)$. Hence
\begin{equation}
h^{\mr{osc}}_{k,N}=\frac{N-1}{8}
+\frac12\nu_{k,N}(1-\nu_{k,N}).
\label{eq:S-hosc}
\end{equation}
This oscillator contribution is independent of the Luttinger parameter $K$.

The principal-angle transformation diagonalizes the local oscillator problem over $\mathbb R$ but need not preserve the compactification lattice. We therefore quantize the compact zero modes directly in the original replica coordinates.

\section{Compact zero modes and $K$ dependence}
\label{sec:S3}

\subsection{Compact branes and intersection}

Let $q_1=(\bOne_N,\bm0_N)$ and $q_2=(\bm0_N,\bOne_N)$ be integral vectors in the $2N$-dimensional replica-label space. The compact folded boundary branes are
\begin{equation}
\mathcal B_C=\{(y,y):y=\alpha q_1+\beta q_2,\ \alpha,\beta\in S^1\},
\qquad
\mathcal B_U=\{(P_{k,N}v,v):v\in(S^1)^{2N}\}.
\label{eq:S-branes}
\end{equation}
A point on $\mathcal B_C$ also belongs to $\mathcal B_U$ if $P_{k,N}y=y$. For $k>0$ this requires $\alpha=\beta$ modulo $2\pi$. Therefore
\begin{equation}
\mathcal B_C\cap\mathcal B_U
=\{\phi_0\,\bOne_{4N}:\phi_0\in\mathbb R/(2\pi\mathbb Z)\}.
\label{eq:S-intersection}
\end{equation}
The mixed strip thus admits a constant classical configuration $X_{\mr{cl}}=X_0$ with $X_0$ on this circle and zero action. In particular,
\begin{equation}
h^{(0)}_{k,N}=0.
\label{eq:S-h0}
\end{equation}
This is already sufficient for the leading exponent in the main text. We next derive the full compact zero mode spectrum.

\subsection{Momentum and stretching sectors}

\textit{Common Neumann momentum.--}
Let $t_0=(\bOne_{2N},\bOne_{2N})$, so $|t_0|^2=4N$. Along the intersection circle $X=\varphi t_0$, the canonically normalized coordinate has effective radius $\sqrt{4N}$ and momentum $p_\ell=\ell/\sqrt{4N}$ with $\ell\in\mathbb Z$. For the normalization in Eq.~\eqref{eq:S-action}, a Neumann zero mode of momentum $p$ contributes $2Kp^2$ to the open-channel conformal weight. Therefore
\begin{equation}
h_N(\ell)=\frac{K\ell^2}{2N}.
\label{eq:S-hN}
\end{equation}

\textit{Common-Dirichlet stretching.--}
Abbreviate $P=P_{k,N}$. The orthogonal complement of the graph $V_U$ is
\begin{equation}
V_U^\perp=\{(r,-Pr):r\in\mathbb R^{2N}\}.
\label{eq:S-VUperp}
\end{equation}
Define the primitive integral vector $a_k=q_1-Pq_1=-(q_2-Pq_2)$, with $a_k^2=2k$. Orthogonality to $V_C$ reduces to $a_k\cdot r=0$, so the common-Dirichlet subspace is
\begin{equation}
W=V_C^\perp\cap V_U^\perp
=\{(r,-Pr):a_k\cdot r=0\},
\qquad \dim W=2N-1.
\label{eq:S-W}
\end{equation}
Lift the toroidal branes to $\mathbb R^{4N}$. Two lifts may differ by $2\pi w$, with $w=(w_1,w_2)$ and $w_1,w_2\in\mathbb Z^{2N}$. For fixed $w$, minimizing over endpoints along the two branes leaves the shortest displacement $\Delta_w=2\pi\Pi_Ww$, where $\Pi_W$ is the orthogonal projector onto $W$. The static harmonic strip configuration is linear across a strip of width $\pi$,
\begin{equation}
X_{\mr{cl}}(\sigma)=X_C+\frac{\sigma}{\pi}\Delta_w,
\qquad 0\le\sigma\le\pi,
\label{eq:S-Xcl}
\end{equation}
and contributes
\begin{equation}
h_D(w)=\frac{1}{8\pi^2K}|\Delta_w|^2
=\frac{1}{2K}|\Pi_Ww|^2.
\label{eq:S-hDw}
\end{equation}
Define $z=w_1-Pw_2\in\mathbb Z^{2N}$. If $\Pi_Ww=(r,-Pr)$, projection onto the hyperplane orthogonal to $a_k$ gives
\begin{equation}
2r=\Pi_{a_k^\perp}z
=z-\frac{a_k(a_k\cdot z)}{2k}.
\label{eq:S-rproj}
\end{equation}
Therefore
\begin{equation}
|\Pi_Ww|^2=\frac12\left[z^2-\frac{(a_k\cdot z)^2}{2k}\right].
\label{eq:S-projectnorm}
\end{equation}
Only the projection of $z$ orthogonal to $a_k$ matters. Since $a_k$ is primitive, the kernel of this projection inside the integer lattice is exactly $\mathbb Za_k$. Distinct stretching sectors are therefore labeled by $[z]\in\mathbb Z^{2N}/\mathbb Za_k$, with classical weight
\begin{equation}
h_D([z])=\frac{1}{4K}
\left[z^2-\frac{(a_k\cdot z)^2}{2k}\right].
\label{eq:S-hDz}
\end{equation}
The expression is invariant under $z\mapsto z+qa_k$ and is nonnegative.

\subsection{Full compact spectrum and ground sector}

Combining the common Neumann momentum with the common Dirichlet stretching sectors yields
\begin{equation}
h^{(0)}_{k,N;\ell,[z]}
=\frac{K\ell^2}{2N}
+\frac{1}{4K}
\left[z^2-\frac{(a_k\cdot z)^2}{2k}\right],
\qquad
\ell\in\mathbb Z,\quad [z]\in\mathbb Z^{2N}/\mathbb Za_k.
\label{eq:S-compactfull}
\end{equation}
The ND and rotated sectors carry no independent zero modes; their effects are already contained in the fractional oscillator spectrum. Thus a general BCC state has
\begin{equation}
h_{\mr{BCC}}=h^{\mr{osc}}_{k,N}+h^{(0)}_{k,N;\ell,[z]}+\Delta h_{\mr{osc}},
\label{eq:S-hBCC}
\end{equation}
where $\Delta h_{\mr{osc}}\ge0$ denotes an oscillator excitation above the BCC vacuum.
The minimum is attained at $\ell=0$ and $[z]=[0]$, for which $h^{(0)}_{k,N;0,[0]}=0$. This is precisely the constant configuration on the common diagonal circle in Eq.~\eqref{eq:S-intersection}.

\section{Frame potential inequalities}
\label{sec:S4}

For the complete rank-one measurement protocol considered here, set
$X\equiv X_{\mout\nout}=|\langle\psi_{\mout}|\psi_{\nout}\rangle|^2$. Then $0\le X\le1$ and $F_k=\mathbb E[X^k]$ under the probability measure $p_{\mout}p_{\nout}$. Standard moment inequalities give
\begin{equation}
1\ge F_k\ge F_{k+1},
\label{eq:S-monotone}
\end{equation}
\begin{equation}
F_k\ge F_1^k,
\label{eq:S-Jensen}
\end{equation}
and the log-convexity relation
\begin{equation}
F_k^2\le F_{k-1}F_{k+1}.
\label{eq:S-logconvex}
\end{equation}

The predicted exponent
\begin{equation}
\Gamma(k)=\frac14+\frac{1}{\pi^2}\arcosh^2\!\sqrt{k},
\qquad k\ge1,
\label{eq:S-Gammafunction}
\end{equation}
is consistent with all three constraints. For the inequality analysis we extend $\Gamma_k$ continuously to real $k\ge1$. Writing $x=\arcosh\sqrt{k}$, so $k=\cosh^2x$, one finds
\begin{equation}
\Gamma'(k)=\frac{2x}{\pi^2\sinh2x}>0,
\qquad
\Gamma''(k)=\frac{2[\sinh2x-2x\cosh2x]}{\pi^2\sinh^3 2x}<0
\quad (x>0).
\label{eq:S-Gammaderivs}
\end{equation}
Thus the predicted asymptotic frame-potential hierarchy is consistent with monotonicity and log-convexity. Moreover,
\begin{equation}
\chi_k\equiv k\Gamma_1-\Gamma_k
=\frac{k-1}{4}-\frac{1}{\pi^2}\arcosh^2\!\sqrt{k}\ge0,
\label{eq:S-chik}
\end{equation}
where the last inequality follows from $k-1=\sinh^2x$ and $\sinh x\ge x\ge2x/\pi$ for $x\ge0$. Hence $\Gamma_k\le k\Gamma_1$, as required by Eq.~\eqref{eq:S-Jensen}.

\section{Numerical methods}
\label{sec:S5}

\subsection{Interacting XXZ MPS calculation}

\textit{Ground states and simulation parameters.--}
We numerically test the continuum prediction in the critical spin-$1/2$ XXZ chain
\begin{equation}
H=\sum_{j=1}^{L}\left(
S_j^xS_{j+1}^x+S_j^yS_{j+1}^y+\Delta S_j^zS_{j+1}^z
\right),
\qquad \bm S_{L+1}\equiv\bm S_1,
\label{eq:S-XXZ}
\end{equation}
with $J_{xy}=1$ and periodic boundary conditions. Ground states are represented as matrix-product states and obtained by DMRG \cite{White1992,White1993} using the ITensor library \cite{FishmanWhiteStoudenmire2022}. Abelian quantum numbers are conserved throughout, and the calculation is restricted to total $S^z=0$. We use 14 DMRG sweeps with truncation cutoff $10^{-10}$ and the maximum bond dimensions listed in Table~\ref{tab:production}. A converged MPS checkpoint is stored once for each $(L,\Delta)$ and reused for independent Monte Carlo seeds.

At zero magnetization the Bethe-ansatz relation is
$K(\Delta)=\pi/[2(\pi-\arccos\Delta)]$ for $-1<\Delta\le1$ \cite{Haldane1981,Giamarchi2003}. The main-text data use $\Delta=-0.8,-0.5,0,0.5,0.8$, all within the critical TLL phase, with $\Delta=0$ corresponding to $K=1$.

\begin{table}[t]
\caption{Production parameters for the XXZ data shown in the main-text figure. Three independent Monte Carlo seeds are used for each $(L,\Delta)$. The number of sampled outcome pairs and batches is quoted per seed.}
\label{tab:production}
\begin{ruledtabular}
\begin{tabular}{cccc}
$L$ & requested $\chi_{\max}$ & outcome pairs & batches\\
\hline
32  & 1200 & 30000 & 30\\
48  & 1200 & 35000 & 35\\
64  & 1200 & 50000 & 50\\
80  & 1200 & 60000 & 60\\
96  & 1200 & 75000 & 75\\
128 & 1600 & 90000 & 90\\
\end{tabular}
\end{ruledtabular}
\end{table}

The unmeasured subsystem is the half chain $U=\{1,\ldots,L/2\}$. In lattice units we set $\epsilon=1$, so $D_U=L/\pi$, and projectively measure the complement $C$ in the local $S^z$ basis.

\textit{Born sampling and conditional-state overlaps.--}
For each converged ground-state MPS, we draw complete spin configurations from the Born distribution using the MPS sampling routine. Only the configuration on the measured region $C$ is retained. Two independent draws therefore generate two independent Born-distributed measurement records $\mout$ and $\nout$.

For a fixed record, the MPS tensors on $C$ are projected onto the corresponding local $S^z$ basis states and contracted from the right. With the MPS orthogonality center placed at the $U|C$ cut, the remaining tensor represents the normalized conditional state on $U$. The squared overlap of two conditional states is $X_{\mout\nout}=|\langle\psi_{\mout}|\psi_{\nout}\rangle|^2$, with $0\le X_{\mout\nout}\le1$. The frame potentials are therefore estimated directly as the Born averages $F_k(U)=\mathbb E_{\mout,\nout}[X_{\mout\nout}^k]$ for $k=2,3$. Because the XXZ ground state has fixed total $S^z$, records that imply different $S^z$ sectors for the unmeasured interval produce orthogonal conditional states. Such pairs have $X_{\mout\nout}=0$ and need no explicit overlap contraction.

\textit{Control variate and statistical errors.--}
For the same bipartition the first moment is known independently from the Schmidt spectrum, $F_1(U)=\Tr\rho_U^2$. We use this exact MPS value as a control variate for the Monte Carlo estimators of $F_2$ and $F_3$. For $Y=X_{\mout\nout}^k$ and $\mu_X=F_1$, the corrected batch estimator is
\begin{equation}
Y_{\mr{cv}}=Y-\beta(X-\mu_X).
\label{eq:S-control}
\end{equation}
Here $X$ and $Y$ denote the batch means. The coefficient $\beta$ is cross-fitted at the batch level: when correcting one batch, $\beta$ is estimated only from samples outside that batch. The standard error for each seed is obtained from the variance of the corrected batch means.

The three independent seeds at fixed $(L,\Delta)$ are combined by inverse variance weighting. If their scatter is larger than expected from the individual standard errors, the final error bar is enlarged by the Birge factor
\begin{equation}
B=\max\!\left(1,\sqrt{\frac{Q}{n_{\mr{rep}}-1}}\right),
\label{eq:S-Birge}
\end{equation}
where $Q$ is the weighted replicate chi-square and $n_{\mr{rep}}=3$. The main text figure contains these combined estimates and their statistical errors.

\subsection{Free-fermion calculation at physical $N=1$}

At $\Delta=0$ the XXZ chain maps to a half-filled free fermion Slater determinant, enabling an independent calculation of the projected ensemble directly at the physical replica number $N=1$. We use system sizes divisible by four and choose the antiperiodic half-filled Fermi sea
\begin{equation}
q_\alpha=\frac{2\pi}{L}\left(n_\alpha+\frac12\right),
\qquad
n_\alpha=-\frac L4,\ldots,\frac L4-1,
\label{eq:S-momenta}
\end{equation}
with occupied-orbital matrix
\begin{equation}
\mathcal{V}_{x\alpha}=\frac{e^{iq_\alpha x}}{\sqrt L},
\qquad
x=0,\ldots,L-1,\quad \alpha=1,\ldots,L/2.
\label{eq:S-Vmatrix}
\end{equation}
For a complete occupation configuration $S$ of $L/2$ fermions, the Born probability is
\begin{equation}
p(S)=|\det \mathcal{V}_S|^2,
\label{eq:S-DPPprob}
\end{equation}
where $\mathcal{V}_S$ denotes the $L/2\times L/2$ submatrix obtained by restricting the rows of $\mathcal{V}$ to the occupied sites $S$. The full configurations form the projection determinantal point process with kernel $\mathcal{V}\mathcal{V}^\dagger$. We sample this process exactly and retain only the occupation record on the measured complement $C$.

For a sampled configuration $S$, let $\mathcal{W}=\mathcal{V}_S^{-1}$ and let $J_U$ denote the set of column indices corresponding to sampled particles lying in $U$. The conditional state on $U$ remains a Slater determinant, which can be constructed efficiently using the free-fermion post-measurement formalism described in Refs. \cite{Bravyi:2004,Chen:2026cel}. Writing $Q_{\mout}=\mathcal{W}_{:,J_U}$ and $H_U=\mathcal{V}_U^\dagger \mathcal{V}_U$, its normalized overlap with a second conditional state is evaluated as
\begin{equation}
X_{\mout\nout}=
\frac{\left|\det\!\left(Q_{\mout}^\dagger H_UQ_{\nout}\right)\right|^2}
{\det\!\left(Q_{\mout}^\dagger H_UQ_{\mout}\right)
 \det\!\left(Q_{\nout}^\dagger H_UQ_{\nout}\right)}.
\label{eq:S-ffoverlap}
\end{equation}
Pairs with different particle numbers in $U$ are orthogonal. We therefore stratify the Monte Carlo sampling by the particle number sector on $C$ and use the exactly known first overlap moment within each sector as a control variate. This substantially reduces the variance of $F_k=\mathbb E[X_{\mout\nout}^k]$ for $k=2,3,4$.

The first moment entering the normalized hierarchy is obtained independently from the free-fermion correlation matrix $C_U$,
\begin{equation}
F_1(L,l)=\Tr\rho_U^2
=\det\!\left[C_U^2+(I-C_U)^2\right].
\label{eq:S-F1ff}
\end{equation}
Define $R_k(L,l)=F_k(L,l)/F_1(L,l)^k$. Normalizing by the half-chain value removes both the nonuniversal amplitude and the overall power of $L/\epsilon$. The parameter-free ratio is
\begin{equation}
\frac{R_k(L,l)}{R_k(L,L/2)}
=\left[\sin\!\left(\frac{\pi l}{L}\right)\right]^{\chi_k}[1+o(1)].
\label{eq:S-shaperatio}
\end{equation}
The production calculation uses $L=200,300,400$, interval fractions $l/L=0.10,0.15,\ldots,0.50$, and $k=2,3,4$. For each size we use 16 statistically independent blocks, with 250 DPP-pair iterations per block; the block-to-block scatter gives the quoted standard errors. Main-text Fig.~2(b) compares the resulting ratios with Eq.~\eqref{eq:S-shaperatio}.


\begin{thebibliography}{99}

\bibitem{HoChoi2022}
W. W. Ho and S. Choi,
Exact Emergent Quantum State Designs from Quantum Chaotic Dynamics,
\href{https://doi.org/10.1103/PhysRevLett.128.060601}{Phys. Rev. Lett. \textbf{128}, 060601 (2022)}.

\bibitem{Cotler2023}
J. S. Cotler, D. K. Mark, H.-Y. Huang, F. Hern\'andez, J. Choi, A. L. Shaw, M. Endres, and S. Choi,
Emergent Quantum State Designs from Individual Many-Body Wave Functions,
\href{https://doi.org/10.1103/PRXQuantum.4.010311}{PRX Quantum \textbf{4}, 010311 (2023)}.

\bibitem{IppolitiHo2023}
M. Ippoliti and W. W. Ho,
Dynamical Purification and the Emergence of Quantum State Designs from the Projected Ensemble,
\href{https://doi.org/10.1103/PRXQuantum.4.030322}{PRX Quantum \textbf{4}, 030322 (2023)}.

\bibitem{IppolitiHo2022}
M. Ippoliti and W. W. Ho,
Solvable Model of Deep Thermalization with Distinct Design Times,
\href{https://doi.org/10.22331/q-2022-12-29-886}{Quantum \textbf{6}, 886 (2022)}.

\bibitem{ClaeysLamacraft2022}
P. W. Claeys and A. Lamacraft,
Emergent Quantum State Designs and Biunitarity in Dual-Unitary Circuit Dynamics,
\href{https://doi.org/10.22331/q-2022-06-15-738}{Quantum \textbf{6}, 738 (2022)}.

\bibitem{Choi2023Nature}
J. Choi, A. L. Shaw, I. S. Madjarov, X. Xie, R. Finkelstein, J. P. Covey, J. S. Cotler, D. K. Mark, H.-Y. Huang, A. Kale, H. Pichler, F. G. S. L. Brand\~ao, S. Choi, and M. Endres,
Preparing Random States and Benchmarking with Many-Body Quantum Chaos,
\href{https://doi.org/10.1038/s41586-022-05442-1}{Nature \textbf{613}, 468--473 (2023)}.

\bibitem{Lucas2023}
M. Lucas, L. Piroli, J. De Nardis, and A. De Luca,
Generalized Deep Thermalization for Free Fermions,
\href{https://doi.org/10.1103/PhysRevA.107.032215}{Phys. Rev. A \textbf{107}, 032215 (2023)}.

\bibitem{Mark2024}
D. K. Mark, F. Surace, A. Elben, A. L. Shaw, J. Choi, G. Refael, M. Endres, and S. Choi,
Maximum Entropy Principle in Deep Thermalization and in Hilbert-Space Ergodicity,
\href{https://doi.org/10.1103/PhysRevX.14.041051}{Phys. Rev. X \textbf{14}, 041051 (2024)}.

\bibitem{ChanDeLuca2024}
A. Chan and A. De Luca,
Projected State Ensemble of a Generic Model of Many-Body Quantum Chaos,
\href{https://doi.org/10.1088/1751-8121/ad7211}{J. Phys. A: Math. Theor. \textbf{57}, 405001 (2024)}.

\bibitem{YuHoKos2025}
X.-H. Yu, W. W. Ho, and P. Kos,
Mixed State Deep Thermalization,
\href{https://doi.org/10.1103/t6zs-3f8k}{Phys. Rev. Lett. \textbf{135}, 260402 (2025)}.

\bibitem{ChangEtAl2025}
R.-A. Chang, H. Shrotriya, W. W. Ho, and M. Ippoliti,
Deep Thermalization under Charge-Conserving Quantum Dynamics,
\href{https://doi.org/10.1103/PRXQuantum.6.020343}{PRX Quantum \textbf{6}, 020343 (2025)}.

\bibitem{Yan2026}
Z. Yan, Z.-Y. Ge, R. Li, Y.-R. Zhang, F. Nori, and Y. Nakamura,
Characterizing Many-Body Dynamics with Projected Ensembles on a Superconducting Quantum Processor,
\href{https://doi.org/10.1126/sciadv.aeb8213}{Sci. Adv. \textbf{12}, eaeb8213 (2026)}.

\bibitem{Stephan2009}
J.-M. St\'ephan, S. Furukawa, G. Misguich, and V. Pasquier,
Shannon and Entanglement Entropies of One- and Two-Dimensional Critical Wave Functions,
\href{https://doi.org/10.1103/PhysRevB.80.184421}{Phys. Rev. B \textbf{80}, 184421 (2009)}.

\bibitem{Zaletel2011}
M. P. Zaletel, J. H. Bardarson, and J. E. Moore,
Logarithmic Terms in Entanglement Entropies of 2D Quantum Critical Points and Shannon Entropies of Spin Chains,
\href{https://doi.org/10.1103/PhysRevLett.107.020402}{Phys. Rev. Lett. \textbf{107}, 020402 (2011)}.

\bibitem{Stephan2011}
J.-M. St\'ephan, G. Misguich, and V. Pasquier,
Phase Transition in the R\'enyi-Shannon Entropy of Luttinger Liquids,
\href{https://doi.org/10.1103/PhysRevB.84.195128}{Phys. Rev. B \textbf{84}, 195128 (2011)}.

\bibitem{AlcarazRajabpour2014}
F. C. Alcaraz and M. A. Rajabpour,
Universal Behavior of the Shannon and R\'enyi Mutual Information of Quantum Critical Chains,
\href{https://doi.org/10.1103/PhysRevB.90.075132}{Phys. Rev. B \textbf{90}, 075132 (2014)}.

\bibitem{Rajabpour2015}
M. A. Rajabpour,
Post-Measurement Bipartite Entanglement Entropy in Conformal Field Theories,
\href{https://doi.org/10.1103/PhysRevB.92.075108}{Phys. Rev. B \textbf{92}, 075108 (2015)}.

\bibitem{Rajabpour2016}
M. A. Rajabpour,
Entanglement Entropy after a Partial Projective Measurement in $1+1$ Dimensional Conformal Field Theories: Exact Results,
\href{https://doi.org/10.1088/1742-5468/2016/06/063109}{J. Stat. Mech. (2016) 063109}.

\bibitem{NajafiRajabpour2016}
K. Najafi and M. A. Rajabpour,
Entanglement Entropy after Selective Measurements in Quantum Chains,
\href{https://doi.org/10.1007/JHEP12(2016)124}{J. High Energy Phys. \textbf{12} (2016) 124}.


\bibitem{Garratt2023}
S. J. Garratt, Z. Weinstein, and E. Altman,
Measurements Conspire Nonlocally to Restructure Critical Quantum States,
\href{https://doi.org/10.1103/PhysRevX.13.021026}{Phys. Rev. X \textbf{13}, 021026 (2023)}.

\bibitem{Weinstein2023}
Z. Weinstein, R. Sajith, E. Altman, and S. J. Garratt,
Nonlocality and Entanglement in Measured Critical Quantum Ising Chains,
\href{https://doi.org/10.1103/PhysRevB.107.245132}{Phys. Rev. B \textbf{107}, 245132 (2023)}.

\bibitem{Yang2023}
Z. Yang, D. Mao, and C.-M. Jian,
Entanglement in a One-Dimensional Critical State after Measurements,
\href{https://doi.org/10.1103/PhysRevB.108.165120}{Phys. Rev. B \textbf{108}, 165120 (2023)}.

\bibitem{Murciano2023}
S. Murciano, P. Sala, Y. Liu, R. S. K. Mong, and J. Alicea,
Measurement-Altered Ising Quantum Criticality,
\href{https://doi.org/10.1103/PhysRevX.13.041042}{Phys. Rev. X \textbf{13}, 041042 (2023)}.

\bibitem{LeeJianXu2023}
J. Y. Lee, C.-M. Jian, and C. Xu,
Quantum Criticality Under Decoherence or Weak Measurement,
\href{https://doi.org/10.1103/PRXQuantum.4.030317}{PRX Quantum \textbf{4}, 030317 (2023)}.

\bibitem{SunYaoJian2023}
X. Sun, H. Yao, and S.-K. Jian,
New Critical States Induced by Measurement,
\href{https://arxiv.org/abs/2301.11337}{arXiv:2301.11337 (2023)}.

\bibitem{Cheng2024}
Z. Cheng, R. Wen, S. Gopalakrishnan, R. Vasseur, and A. C. Potter,
Universal Structure of Measurement-Induced Information in Many-Body Ground States,
\href{https://doi.org/10.1103/PhysRevB.109.195128}{Phys. Rev. B \textbf{109}, 195128 (2024)}.

\bibitem{KhannaMurcianoVasseur2026}
K. Khanna, S. Murciano, and R. Vasseur,
Theory of Measurement-Altered Criticality,
\href{https://arxiv.org/abs/2608.05289}{arXiv:2608.05289 (2026)}.

\bibitem{KumarPatilLudwigVasseur2026}
A. Kumar, R. A. Patil, A. W. W. Ludwig, and R. Vasseur,
Universal Crossovers in Weakly-Monitored Quantum Critical States,
\href{https://arxiv.org/abs/2608.02716}{arXiv:2608.02716 (2026)}.

\bibitem{HoshinoOshikawaAshida2025}
M. Hoshino, M. Oshikawa, and Y. Ashida,
Entanglement Swapping in Critical Quantum Spin Chains,
\href{https://doi.org/10.1103/PhysRevB.111.155143}{Phys. Rev. B \textbf{111}, 155143 (2025)}.

\bibitem{Khanna:2025mrz}
K.~Khanna and R.~Vasseur,
Universal Statistics of Measurement-Induced Entanglement in Tomonaga-Luttinger liquids,
\href{https://arxiv.org/abs/2512.13809}{arXiv:2512.13809 (2025)}.

\bibitem{KhannaVasseur2026}
K. Khanna and R. Vasseur,
Measurement-Induced Entanglement in Conformal Field Theory,
\href{https://doi.org/10.1103/b7sb-nhjq}{Phys. Rev. Lett. \textbf{136}, 160402 (2026)}.



\bibitem{MilekhinMurciano2025}
A. Milekhin and S. Murciano,
Observable-Projected Ensembles,
\href{https://doi.org/10.22331/q-2025-10-20-1888}{Quantum \textbf{9}, 1888 (2025)}.

\bibitem{Haldane1981}
F. D. M. Haldane,
`Luttinger Liquid Theory' of One-Dimensional Quantum Fluids. I. Properties of the Luttinger Model and Their Extension to the General 1D Interacting Spinless Fermi Gas,
\href{https://doi.org/10.1088/0022-3719/14/19/010}{J. Phys. C: Solid State Phys. \textbf{14}, 2585 (1981)}.

\bibitem{Giamarchi2003}
T. Giamarchi,
\emph{Quantum Physics in One Dimension} (Oxford University Press, 2003).

\bibitem{Cardy1984}
J. L. Cardy,
Conformal Invariance and Surface Critical Behavior,
\href{https://doi.org/10.1016/0550-3213(84)90241-4}{Nucl. Phys. B \textbf{240}, 514 (1984)}.

\bibitem{Cardy1989}
J. L. Cardy,
Boundary Conditions, Fusion Rules and the Verlinde Formula,
\href{https://doi.org/10.1016/0550-3213(89)90521-X}{Nucl. Phys. B \textbf{324}, 581 (1989)}.

\bibitem{GaberdielRecknagel2001}
M. R. Gaberdiel and A. Recknagel,
Conformal Boundary States for Free Bosons and Fermions,
\href{https://doi.org/10.1088/1126-6708/2001/11/016}{J. High Energy Phys. \textbf{11} (2001) 016}.

\bibitem{BjorckGolub1973}
\AA. Bj\"orck and G. H. Golub,
Numerical Methods for Computing Angles Between Linear Subspaces,
\href{https://doi.org/10.1090/S0025-5718-1973-0348991-3}{Math. Comp. \textbf{27}, 579 (1973)}.

\bibitem{BerkoozDouglasLeigh1996}
M. Berkooz, M. R. Douglas, and R. G. Leigh,
Branes Intersecting at Angles,
\href{https://doi.org/10.1016/S0550-3213(96)00452-X}{Nucl. Phys. B \textbf{480}, 265 (1996)}.

\bibitem{Pesando2014}
I. Pesando,
Canonical Quantization of a String Describing $N$ Branes at Angles,
\href{https://doi.org/10.1016/j.nuclphysb.2014.10.005}{Nucl. Phys. B \textbf{889}, 120 (2014)}.



\bibitem{SupplementalMaterial}
See Supplemental Material 
for details of the auxiliary-replica construction, folded boundary CFT,
compact zero modes, moment inequalities, and numerical methods.

\bibitem{CalabreseCardy2004}
P. Calabrese and J. Cardy,
Entanglement Entropy and Quantum Field Theory,
\href{https://doi.org/10.1088/1742-5468/2004/06/P06002}{J. Stat. Mech. (2004) P06002}.

\bibitem{White1992}
S. R. White,
Density Matrix Formulation for Quantum Renormalization Groups,
\href{https://doi.org/10.1103/PhysRevLett.69.2863}{Phys. Rev. Lett. \textbf{69}, 2863 (1992)}.

\bibitem{White1993}
S. R. White,
Density-Matrix Algorithms for Quantum Renormalization Groups,
\href{https://doi.org/10.1103/PhysRevB.48.10345}{Phys. Rev. B \textbf{48}, 10345 (1993)}.

\bibitem{ITensor2022}
M. Fishman, S. R. White, and E. M. Stoudenmire,
The ITensor Software Library for Tensor Network Calculations,
\href{https://doi.org/10.21468/SciPostPhysCodeb.4}{SciPost Phys. Codebases \textbf{4} (2022)}.

\bibitem{GoldsteinSela2018}
M. Goldstein and E. Sela,
Symmetry-Resolved Entanglement in Many-Body Systems,
\href{https://doi.org/10.1103/PhysRevLett.120.200602}{Phys. Rev. Lett. \textbf{120}, 200602 (2018)}.

\bibitem{BonsignoriRuggieroCalabrese2019}
R. Bonsignori, P. Ruggiero, and P. Calabrese,
Symmetry Resolved Entanglement in Free Fermionic Systems,
\href{https://doi.org/10.1088/1751-8121/ab4b77}{J. Phys. A: Math. Theor. \textbf{52}, 475302 (2019)}.

\bibitem{KusukiMurcianoOoguriPal2023}
Y. Kusuki, S. Murciano, H. Ooguri, and S. Pal,
Symmetry-Resolved Entanglement Entropy, Spectra, and Boundary Conformal Field Theory,
\href{https://doi.org/10.1007/JHEP11(2023)216}{J. High Energy Phys. \textbf{11} (2023) 216}.

\bibitem{AntoniniEtAl2022}
S. Antonini, G. Bentsen, C. Cao, J. Harper, S.-K. Jian, and B. Swingle,
Holographic Measurement and Bulk Teleportation,
\href{https://doi.org/10.1007/JHEP12(2022)124}{J. High Energy Phys. \textbf{12} (2022) 124}.

\bibitem{AntoniniEtAl2023}
S. Antonini, B. Grado-White, S.-K. Jian, and B. Swingle,
Holographic Measurement in CFT Thermofield Doubles,
\href{https://doi.org/10.1007/JHEP07(2023)014}{J. High Energy Phys. \textbf{07} (2023) 014}.


\end{thebibliography}

\begin{thebibliography}{14}

\bibitem{Haldane1981}
F. D. M. Haldane,
`Luttinger Liquid Theory' of One-Dimensional Quantum Fluids. I. Properties of the Luttinger Model and Their Extension to the General 1D Interacting Spinless Fermi Gas,
\href{https://doi.org/10.1088/0022-3719/14/19/010}{J. Phys. C: Solid State Phys. \textbf{14}, 2585 (1981)}.

\bibitem{Giamarchi2003}
T. Giamarchi,
\emph{Quantum Physics in One Dimension} (Oxford University Press, Oxford, 2003).


\bibitem{Cardy1989}
J. L. Cardy,
Boundary Conditions, Fusion Rules and the Verlinde Formula,
\href{https://doi.org/10.1016/0550-3213(89)90521-X}{Nucl. Phys. B \textbf{324}, 581 (1989)}.

\bibitem{GaberdielRecknagel2001}
M. R. Gaberdiel and A. Recknagel,
Conformal Boundary States for Free Bosons and Fermions,
\href{https://doi.org/10.1088/1126-6708/2001/11/016}{J. High Energy Phys. \textbf{11} (2001) 016}.

\bibitem{BerkoozDouglasLeigh1996}
M. Berkooz, M. R. Douglas, and R. G. Leigh,
Branes Intersecting at Angles,
\href{https://doi.org/10.1016/S0550-3213(96)00452-X}{Nucl. Phys. B \textbf{480}, 265 (1996)}.

\bibitem{Pesando2014}
I. Pesando,
Canonical Quantization of a String Describing $N$ Branes at Angles,
\href{https://doi.org/10.1016/j.nuclphysb.2014.10.005}{Nucl. Phys. B \textbf{889}, 120 (2014)}.

\bibitem{BjorckGolub1973}
\AA. Bj\"orck and G. H. Golub,
Numerical Methods for Computing Angles Between Linear Subspaces,
\href{https://doi.org/10.1090/S0025-5718-1973-0348991-3}{Math. Comp. \textbf{27}, 579 (1973)}.

\bibitem{White1992}
S. R. White,
Density Matrix Formulation for Quantum Renormalization Groups,
\href{https://doi.org/10.1103/PhysRevLett.69.2863}{Phys. Rev. Lett. \textbf{69}, 2863 (1992)}.

\bibitem{White1993}
S. R. White,
Density-Matrix Algorithms for Quantum Renormalization Groups,
\href{https://doi.org/10.1103/PhysRevB.48.10345}{Phys. Rev. B \textbf{48}, 10345 (1993)}.

\bibitem{FishmanWhiteStoudenmire2022}
M. Fishman, S. R. White, and E. M. Stoudenmire,
The ITensor Software Library for Tensor Network Calculations,
\href{https://doi.org/10.21468/SciPostPhysCodeb.4}{SciPost Phys. Codebases \textbf{4} (2022)}.

\bibitem{Bravyi:2004}
S.~Bravyi,
``Lagrangian representation for fermionic linear optics,''
\href{https://arxiv.org/abs/quant-ph/0404180}{Quantum Inf. Comput. \textbf{5}, 216--238 (2005)}.

\bibitem{Chen:2026cel}
H.-H.~Chen,
``Entanglement of excited states after measurements in conformal field theory,''
\href{https://arxiv.org/abs/2607.04268}{arXiv:2607.04268 (2026)}.
\end{thebibliography}
\end{document}